\documentclass[twocolumn,showpacs,preprintnumbers,amsmath,amssymb]{revtex4}
\usepackage{graphicx}
\usepackage{dcolumn}
\usepackage{bm}

\begin{document}

\preprint{APS/123-QED}
\title{Feshbach resonances in mixtures of ultracold $^6$Li and $^{87}$Rb gases}

\author{B. Deh, C. Marzok, C. Zimmermann, and Ph.W. Courteille}
\affiliation{Physikalisches Institut, Eberhard-Karls-Universit\"at T\"ubingen,
\\Auf der Morgenstelle 14, D-72076 T\"ubingen, Germany}

\date{\today}

\begin{abstract}
We report on the observation of two Feshbach resonances in collisions between ultracold $^6$Li and $^{87}$Rb atoms in their respective hyperfine ground states $|F,m_F\rangle=|1/2,1/2\rangle$ and $|1,1\rangle$. The resonances show up as trap losses for the $^6$Li cloud induced by inelastic Li-Rb-Rb three-body collisions. The magnetic field values where they occur represent important benchmarks for an accurate determination of the interspecies interaction potentials. A broad Feshbach resonance located at 1066.92~G opens interesting prospects for the creation of ultracold heteronuclear molecules. We furthermore observe a strong enhancement of the narrow $p$-wave Feshbach resonance in collisions of $^6$Li atoms at 158.55~G in the presence of a dense $^{87}$Rb cloud. The effect of the $^{87}$Rb cloud is to introduce Li-Li-Rb three-body collisions occurring at a higher rate than Li-Li-Li collisions.
\end{abstract}


\maketitle


The coherent and superfluid dynamics of ultracold atomic gases is essentially ruled by two parameters, the atomic mass and the $s$-wave scattering length for interatomic collisions. The scattering length can be controlled by applying a magnetic field whose strength is chosen to address a Feshbach collision resonance \cite{Inouye98}. Since the location of Feshbach resonances is very sensitive to the shape of the long-range part of the interatomic potentials, their experimental observation sets tight constraints to them and determines the scattering lengths. This also holds when the gas under consideration is a mixture of different atomic species. For several years research is being performed on mixtures of $^6$Li-$^7$Li \cite{Truscott01}, $^{85}$Rb-$^{87}$Rb \cite{Bloch01,Papp06}, $^{40}$K-$^{87}$Rb \cite{Roati02,Inouye04,OspelkausC06}, and $^6$Li-$^{23}$Na \cite{Hadzibabic02}. The identification of interspecies Feshbach resonances \cite{Stan04,Inouye04} led to a detailed knowledge of the interspecies scattering potentials. Unfortunately, for collisions between Li and Rb the interaction potentials are unknown to the point that even the sign of the zero magnetic field scattering length remains unidentified \cite{Silber05,Marzok07}. These problems can now be resolved due to the experimental determination of the locations of two Feshbach resonances reported in this paper.


The present interest in mixtures of ultracold gases has many motivations. To begin with a practical application, mixtures allow for sympathetic cooling of one species via another. This is particularly important for cooling pure Fermi gases, which cannot be cooled actively. 
While the $^6$Li-$^{87}$Rb combination appears to be particularly well suited for reaching low temperatures \cite{Brown-Hayes04,Cote05}, the smallness of the interspecies $s$-wave scattering length slows down the sympathetic cooling rate \cite{Silber05}. This could be a problem if heating processes such as Fermi hole heating become important \cite{Timmermans01,Cote05}.
 

At very low temperatures boson-mediated Cooper pairing is predicted, a type of fermionic superfluidity bearing similarities with the phenomenon of superconductivity in metals \cite{Heiselberg00}. Differently from Feshbach resonance-induced condensation of fermions \cite{Herbig03}, this kind of Cooper pairing relies on the presence of bosons mediating the fermionic interaction. Interspecies mean-field interaction has already been observed in $^{40}$K-$^{87}$Rb mixtures, where it lead to collapsing atomic clouds \cite{OspelkausC06,Modugno02}.


If bound in low-lying vibrational states LiRb molecules exhibit a particularly large electric dipole moment \cite{Aymar05}. With the recent creation of ultracold heteronuclear dimers near a Feshbach resonance \cite{OspelkausC06b,Papp06,Zirbel07} polar molecular quantum gases become conceivable. Polar gases are fascinating objects expected to show novel global features due to long-ranged intermolecular interactions \cite{Santos00}.


A striking particularity of the LiRb combination is the large mass ratio. The ratio raises the question about the applicability of the Born-Oppenheimer approximation to the intermolecular interaction: The heavy Rb atoms of the LiRb molecules may interact via an exchange of the light Li atoms. Petrov et al. predicted for example the formation of crystalline phases in molecular gases of fermionic constituents, such as $^6$Li-$^{40}$K \cite{Petrov07}. Since the exchange interaction scales with the inverse of the mass of the light atoms, LiRb may be a suitable candidate for studying gaseous solid state physics.


In a previous paper we reported cooling of a mixture of $^6$Li and $^{87}$Rb atoms to simultaneous quantum degeneracy in a magnetic trap \cite{Silber05}. The absolute value of the interspecies $s$-wave scattering length has been determined and found to be small, $|a_s|\simeq20\,a_B$. The present paper presents an extension of this experiment. The mixture is now loaded into a crossed beam optical dipole trap \cite{Adams95}, and a variable homogeneous magnetic field is applied in order to drive the atoms into a Feshbach collision resonance of the scattering cross section. We found two such resonances in the respective ground states of the $^{87}$Rb and the $^6$Li atoms, $|1,1\rangle$ and $|1/2,1/2\rangle$. A broad resonance located at a magnetic field of 1066.92~G may prove useful for generating heteronuclear LiRb molecules.

A strong enhancement of the rate for inelastic processes observed at 158.55~G is due to a $^6$Li $p$-wave Feshbach resonance. In Ref.~\cite{ZhangJ04} the resonance has been found via a resonant enhancement of the rate for Li-Li-Li three-body collisions. Under the conditions of our experiment the weak resonance is barely visible. In contrast, the presence of a dense cloud of $^{87}$Rb atoms catalyses the inelastic processes by promoting the rate for Li-Li-Rb collisions. This demonstrates that it is possible to improve the detection efficiency of weak resonances in dilute sample by providing a dense sample of collision partners.


An experimental run starts with the sequence described in Ref.~\cite{Silber05}. In short, $^6$Li and $^{87}$Rb atoms are simultaneously collected by a magneto-optical trap and then transferred via several intermediate magnetic traps into a Ioffe-Pritchard type trap characterized by the secular frequencies $\omega/2\pi=(50\times206\times206)^{1/3}~$Hz and the magnetic field offset $3.5~$G. Here, the rubidium cloud is selectively cooled by microwave-induced forced evaporation. The $^6$Li cloud adjusts its temperature to the $4~\mu$K cold $^{87}$Rb cloud through interspecies thermalization. The clouds are now slightly decompressed reaching temperatures around $1~\mu$K before being transferred to a dipole trap.

The light for the dipole trap is generated by an Yb-YAG fiber laser (YLD-10-LP, IGP Photonics) operating at 1064~nm. The light is divided into two horizontal laser beams each one having a power of 2.2~W and crossing at right angle at the center of the magnetic trap. Both beams are focused to a beam waist of $w_0=58~\mu$m. For $^{87}$Rb the resulting optical trap is $130~\mu$K deep and has the secular frequencies $\omega_{Rb}/2\pi=(600\times425\times425)^{1/3}~$Hz. For $^6$Li the trap is $52~\mu$K deep and has the secular frequencies $\omega_{Li}/2\pi=(1480\times1050\times1050)^{1/3}~$Hz. The trap frequencies are measured by displacing the atomic clouds in the dipole trap via a magnetic field gradient and monitoring its oscillatory motion once the gradient has been switched off. The clouds thermalize in the dipole trap at temperatures of $7~\mu$K for $^{87}$Rb and $4.8~\mu$K for $^6$Li. The $^6$Li cloud is colder because of additional evaporation in the shallower trap. With typically $4\times10^6$ Rb and $1.4\times10^5$ $^6$Li atoms, the respective densities are $2\times10^{13}~\text{cm}^{-3}$ for $^{87}$Rb and $2\times10^{11}~\text{cm}^{-3}$ for Li. This corresponds to the critical temperature $T_c=3.4~\mu$K and the Fermi-temperature $T_F=5.3~\mu$K.


A microwave field sweep now induces an adiabatic rapid passage of the $^{87}$Rb cloud from the $|2,2\rangle$ to the lowest Zeeman substate $|1,1\rangle$. The sweep is performed in the presence of a homogeneous 4~G magnetic field and covers a range of 3~MHz in 4~ms. We obtain about 98\% transfer efficiency. All Rb atoms remaining in the $|2,2\rangle$ state are removed from the trap by a resonant light pulse. Shortly thereafter the $^6$Li cloud is transferred from the state $|3/2,3/2\rangle$ to the lowest Zeeman substate $|1/2,1/2\rangle$ using the same technique with a radiofrequency sweep of 3~ms duration and a span of 3.2~MHz. Now the homogeneous magnetic field is quickly increased with a fast ramping speed between 20 and 73~G/ms to a variable final value $B$ \cite{Note2}, where it is held for a time interval chosen between $\Delta t=5~$ms and $200~$ms. The interaction dynamics studied in this paper takes place within this interval. Finally, the magnetic field is switched off suddenly, the dipole trap is shut down, and both clouds are absorption-imaged after a free expansion time of 1~ms for $^6$Li and 5~ms for $^{87}$Rb. The main information extracted from the images is the number of $^6$Li and $^{87}$Rb atoms remaining in the trap as a function of the magnetic field strength $B$ and the interaction time $\Delta t$.

The unresolved hyperfine structure of lithium hinder a precise measurement of atom numbers through absorption imaging at vanishing magnetic fields \cite{Marzok07}. We circumvent this problem by calibrating the absorption imaging with measurements performed at high magnetic fields, where the Paschen-Back regime ensures the existence of cycling optical transitions.

	\begin{figure}[ht]
		\centerline{\scalebox{.42}{\includegraphics{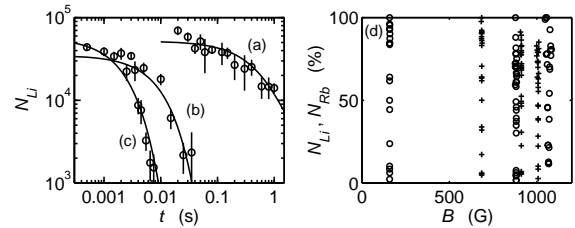}}}
		\caption{(a) $^6$Li decay curve recorded at $B=4~$G far from any Feshbach resonance in the presence of a dense $^{87}$Rb cloud
			of $10^6$ atoms. (b,c) $^6$Li decay curves on the interspecies Feshbach resonances at (b) 882.02~G and (c) 1066.92~G.
			(d) Trap loss spectra near Feshbach resonances for $^6$Li and $^{87}$Rb atoms in their lowest hyperfines states.
			On the scale of this diagram the resonances are not spectrally resolved. The atom numbers remaining in the trap after an
			interaction time $\Delta t$ are normalized to the numbers measured for $\Delta t=0$.
			Crosses (+) correspond to increased losses in the $^{87}$Rb cloud. The three resonances found in this channel are caused 
			by $s$-wave Feshbach resonances in collisions between two $^{87}$Rb atoms \cite{Marte02}. The measured values are listed in
			Tab.~\ref{TabI}. 
			Circles (o) correspond to increased losses in the $^6$Li cloud. The loss feature at 158.55~G is due to a $p$-wave Feshbach 
			resonance in $^6$Li collisions. The remaining two features are due to interspecies Feshbach resonances.
			For those resonances, losses of Rb atoms are not discernible due to their large number overwhelming the Li cloud. }
		\label{Fig1}
	\end{figure}


Inelastic collisions give rise to trap losses. The dominant inelastic processes are three-body collisions, since in the lowest hyperfine states dipolar spinflip is impossible. Collisions with atoms from the background gas only induce negligible decay rates. For the conditions specified above, the measured lifetime of pure $^{87}$Rb or pure $^6$Li clouds (the latter ones are obtained by removing the $^{87}$Rb atoms from the trap by a resonant laser light pulse) lies well beyond 30~s. This is compatible with decay rates on the order of about $K_{Rb,Rb,Rb},K_{Li,Li,Li}\simeq10^{-28}~\text{cm}^6/\text{s}$ \cite{Soeding99}.

The situation changes for a mixture of both species. In the presence of a large $^{87}$Rb cloud, the $^6$Li cloud decays within a few seconds, as can be seen from Fig.~\ref{Fig1}(a), while the $^{87}$Rb cloud is nearly unaffected. The fast decay of the $^6$Li cloud is due to the emergence of new three-body decay channels. Processes involving two $^6$Li and one $^{87}$Rb atoms described by the three-body rate coefficient $K_{Li,Li,Rb}$ are suppressed by the Pauli principle. In contrast, collisions of one $^6$Li and two $^{87}$Rb atoms are possible. Hence, we can set up a simple two-species rate equation model and adjust the rate coefficient $K_{Li,Rb,Rb}$ to the measured decay curve (solid lines in Fig.~\ref{Fig1}). We obtain $K_{{Li,Rb,Rb}}=(6.7\text{-}44)\times10^{-29}~\text{cm}^6/\text{s}$. However we will see below, that the rate coefficient $K_{Li,Li,Rb}$ cannot always be neglected and may even dominate under certain circumstances.
	\begin{figure}[h]
		\centerline{\scalebox{.42}{\includegraphics{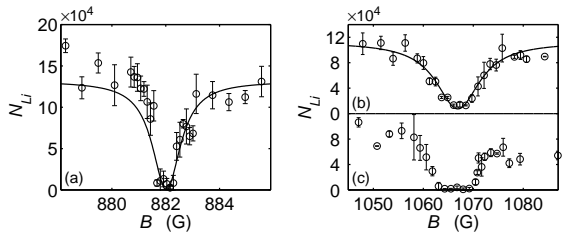}}}
		\caption{(a) Zoom of the trap loss spectrum near 882.02~G for an interaction time $\Delta t=300~$ms.
			(b) Zoom of the trap loss spectrum near 1066.92~G for a short interaction time of 5~ms.
			(c) For a longer interaction time of 30~ms the trap loss saturates.}
		\label{Fig2}
	\end{figure}


The vicinity of a Feshbach resonance reveals itself by a dramatic increase of the scattering length and consequently an increase of the cross section for inelastic collisions [see Fig.~\ref{Fig1}(b,c)]. Interspecies Feshbach resonances will specifically enhance the rate coefficients $K_{Li,Rb,Rb}$. The rate coefficient for the dominating three-body recombination process can again be extracted from lifetime measurements. For the Feshbach resonances located at magnetic fields of 882.02~G and 1066.92~G, we find respectively  $K_{Li,Rb,Rb}^{(res)}=(1.1\text{-}6.1)\times10^{-26}~\text{cm}^6/\text{s}$ and $K_{Li,Rb,Rb}^{(res)}=(7.9\text{-}67)\times10^{-26}~\text{cm}^6/\text{s}$. The main sources of errors for the determination of the rate coefficients are uncertainties in the precise alignment of the two beams forming the crossed dipole trap resulting in uncertain estimates of the local densities. Further systematic corrections may result from errors in the measurement of the atom number and the temperature of the $^6$Li cloud.

The kinetic energy gained from the molecular binding energy in such a collision process is by far too weak to result in trap losses. The observed trap loss processes are due to collision-induced rapid decay of the molecules into deeply bound vibrational states releasing a great amount of kinetic energy. Anomalous decay due to formation of molecular condensates \cite{Abeelen99,Stan04} is not expected for thermal clouds.


To find Feshbach resonances, we searched the magnetic field range between 0~G and 1200~G for increased trap losses. The range was divided into intervals of 3.17~G, and the intervals were swept in subsequent experimental runs within 200~ms. Intervals with increased trap losses were analyzed in more detail. The measured numbers of $^6$Li atoms are plotted as a function of the magnetic field in Figs.~\ref{Fig1}(d) and \ref{Fig2}(a-c). The magnetic field is calibrated by resonant excitation of the $^{87}$Rb hyperfine transition $|1,1\rangle\rightarrow|2,2\rangle$ with a known microwave frequency. Also shown in Fig.~\ref{Fig1}(d) are Rb trap loss spectra near Feshbach resonances in collisions between two $^{87}$Rb atoms, which have been studied earlier \cite{Marte02}.

Several Feshbach resonances have been found at locations listed in Table~\ref{TabI}. Other narrow resonances that may exist within the scanned magnetic field range are below our resolution limit. The profiles of the two interspecies Feshbach resonances have been fitted with Lorentzian functions shown as solid lines in Figs.~\ref{Fig2}(a,b). The fit yields their location with an uncertainty of 0.27~G, limited by systematic errors. It also yields the widths of the resonances exhibited in Tab.~\ref{TabI}. The broadest resonance saturates at interaction times much longer than 5~ms, as shown in Fig.~\ref{Fig2}(c).
	\begin{table}[h]
		\caption{List of detected Feshbach resonances and full width at half maximum of the trap loss features. Also listed is the partial 
			wave $\ell$ of the collision process, the vibrational level $v$ counting from the dissociation limit, and the spin quantum numbers 
			identifying the bound channel \cite{Marte02,ZhangJ04}.}
		\begin{tabular}[c]{c|r|c|c}\hline
			open channel & $B~(\text{G})$ & $\Delta B~(\text{G})$ & $v,\ell(S,I)F,m_F$\\\hline\hline
			$^6\text{Li}|\frac{1}{2},\frac{1}{2}\rangle~^6\text{Li}|\frac{1}{2},\frac{1}{2}\rangle$ & 158.55 & 0.37 & $-1,p(0,1)1,1$\\\textbf{}
			$^{87}\text{Rb}|1,1\rangle~^{87}\text{Rb}|1,1\rangle$ & 684.90 & 0.12 & $-4,s(1,3)\text{-},2$\\
			$^{87}\text{Rb}|1,1\rangle~^6\text{Li}|\frac{1}{2},\frac{1}{2}\rangle$ & 882.02	& 1.27 & ?\\
			$^{87}\text{Rb}|1,1\rangle~^{87}\text{Rb}|1,1\rangle$ & 911.77 & 0.045 & $-5,s(1,3)4,2$\\
			$^{87}\text{Rb}|1,1\rangle~^{87}\text{Rb}|1,1\rangle$ & 1007.58 & 0.29 & $-5,s(1,1)2,2$\\
			$^{87}\text{Rb}|1,1\rangle~^6\text{Li}|\frac{1}{2},\frac{1}{2}\rangle$	& 1066.92 & 10.62 & ?\\\hline
		\end{tabular}
		\label{TabI}
	\end{table}


We searched several other collision channels without finding further interspecies Feshbach resonances in the range of 0~G to 1200~G. Those are the channels $^{87}\text{Rb}|1,0\rangle~^6\text{Li}|1/2,1/2\rangle$, $^{87}\text{Rb}|1,-1\rangle~^6\text{Li}|1/2,1/2\rangle$, and $^{87}\text{Rb}|1,1\rangle~^6\text{Li}|3/2,3/2\rangle$. Channels with $^{87}$Rb in the upper hyperfine state turned out to be too fragile with respect to inelastic collisions.



\begin{figure}[ht]
		\centerline{\scalebox{.42}{\includegraphics{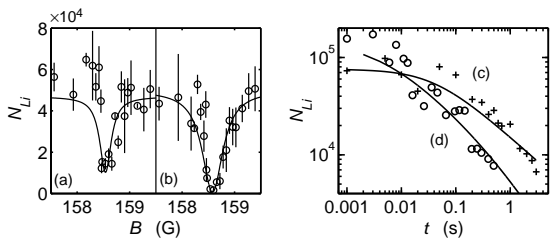}}}
		\caption{$^6$Li trap loss spectra near the $p$-wave Feshbach resonance at 158.55~G in the absence (a) and the presence (b) of a 
			dense $^{87}$Rb cloud. (c) Decay of $^6$Li when $^{87}$Rb is absent from the trap, (d) and decay of $^6$Li when $^{87}$Rb 
			is present in the trap.}
		\label{Fig3}
	\end{figure}

The narrow $^6$Li cloud trap loss feature observed at 158.55~G is due to a $p$-wave Feshbach resonance in collisions between two $^6$Li atoms \cite{ZhangJ04}. In a pure ultracold $^6$Li cloud trap losses due to inelastic Li-Li-Li three-body processes are frozen out behind the centrifugal barrier for $p$-wave collisions. Even when the collision cross section is increased by a Feshbach resonance, enhanced trap losses are barely detectable, in particular when the $^6$Li density is small. Consequently, the loss profile exhibited in Fig.~\ref{Fig3}(a) has a poor signal-to-noise ratio. In contrast we observe a much better signal-to-noise ratio if a $^{87}$Rb cloud is simultaneously trapped [see Fig.~\ref{Fig3}(b)]. This is due to the fact that the Feshbach resonance not only enhances the three-body decay coefficient $K_{Li,Li,Li}$, but also $K_{Li,Li,Rb}$. Applying our rate equation model, at the Feshbach resonance we find $K_{Li,Li,Li}^{(p)}=(2.4\text{-}20)\times10^{-24}~\text{cm}^6/\text{s}$, which is agreement with Ref.~\cite{ZhangJ04}, and $K_{Li,Li,Rb}^{(p)}=(4.7\text{-}95)\times10^{-26}~\text{cm}^6/\text{s}$. Hence, a large $^{87}$Rb cloud can be used as an amplifier for signatures of weak collision resonances, an observation which could be of general importance.


In conclusion, we detected Feshbach collision resonances in a two species ultracold mixture of $^6$Li and $^{87}$Rb gases. The data provide crucial information for the determination of the interspecies interaction potential, which up to now is completely unknown. We anticipate that theoretical calculations based on this interaction potential will help to pin down the interaction potential also for a mixture of bosonic $^7$Li and $^{87}$Rb atoms. With knowledge of the sign of the interspecies triplet scattering length and the location of a suitable Feshbach resonance one could envisage to search for effects such as the interspecies mean-field stabilization of the collapse of a $^7$Li cloud as predicted in Ref.~\cite{Marzok07}.


A goal for future research could be the formation of heteronuclear molecules. Such molecules could be generated by either adiabatically sweeping the magnetic field across an interspecies Feshbach resonance \cite{Papp06}, or by radiofrequency association \cite{OspelkausC06b}. Unfortunately, the molecules are very fragile with respect to collisions with other particles. Supposing the case that all $^6$Li atoms are weakly bound to $^{87}$Rb atoms thus forming fragile molecules, the individual atoms of the molecule will collide with unbound $^{87}$Rb atoms. With the atomic densities specified above, the collision rates are $\gamma_{Rb_2,Rb}\simeq4000~\text{s}^{-1}$ and $\gamma_{Li_2,Rb}\simeq70~\text{s}^{-1}$. The difference in the collision rates is due to the different scattering length for Rb-Rb and for Li-Rb collisions. Hence, the molecular lifetime will be limited by collisions between $^{87}$Rb atoms bound into molecules and free $^{87}$Rb atoms to $\tau\simeq0.25~$ms. To efficiently form heteronuclear molecules, it is thus essential to remove all unbound $^{87}$Rb atoms from the trap. This is confirmed in a recent paper demonstrating that fermionic heteronuclear molecules with large binding energies confined in a single beam dipole trap are surprisingly stable against intermolecular collisions, but very sensitive to the presence of residual unbound $^{87}$Rb atoms \cite{Zirbel07}.



We acknowledge helpful discussions with Alejandro Saenz and Yulian Vanne. This work has been supported by the Deutsche Forschungsgemeinschaft (DFG).

\end{document}